\documentclass[10pt,reqno]{article}
\usepackage{longtable,latexsym,epsfig,caption,subcaption,microtype,textcomp,stmaryrd,listings,lineno,hyperref}

\usepackage{amssymb}
\usepackage{graphicx}
\usepackage{amscd}
\usepackage{amsthm}

\usepackage{float}
\usepackage{graphics, amsmath}
\usepackage{amsfonts}
\usepackage{latexsym}
\usepackage{epsf}
\usepackage{xcolor}
\begin{document}

\theoremstyle{plain}
\newtheorem{theorem}{Theorem}
\newtheorem{corollary}[theorem]{Corollary}
\newtheorem{lemma}{Lemma}
\newtheorem{proposition}[theorem]{Proposition}

\theoremstyle{definition}
\newtheorem{definition}[theorem]{Definition}
\newtheorem{example}[theorem]{Example}
\newtheorem{conjecture}[theorem]{Conjecture}

\theoremstyle{remark}
\newtheorem{remark}{Remark}
\begin{center}
{\Large{{\bf Logistic wavelets and logistic function: An application to model the spread of SARS-CoV-2 virus infections }}}
\medskip
\end{center}
\leftline{}
\leftline{\bf Grzegorz Rz\c{a}dkowski} 

\leftline{Warsaw University of Technology, str. Narbutta 85, 02-524 Warsaw, Poland}
\leftline{e-mail: grzegorz.rzadkowski@pw.edu.pl}

\newcommand{\Eulerian}[2]{\genfrac{<}{>}{0pt}{}{#1}{#2}}

\begin{abstract}
 In the present paper, we model the cumulative number of persons reported to be infected by the SARS-CoV-2 virus, in a country or a region, by a sum of logistic functions. For a given logistic function, using Eulerian numbers, we find the zeros of its successive derivatives and their relationship with the saturation level of this function. In a given time series, having potentially the logistic trend, we use its second differences to determine points corresponding to these zeros. To estimate the parameters of the approximating logistic function, we define and use logistic wavelets. Then we apply the theory to the cases of SARS-CoV-2 infections in the United States and the United Kingdom. 
\end{abstract}



\section{Introduction}

The mathematical modeling of epidemics has a long history; it began with the Kermack-McKendrick model \cite{KM}, introduced in 1927. In this seminal paper, the whole population is divided into Susceptible, Infectious, and Recovered sub-populations. Then,
some ordinary differential equations are formulated specifying the time evolution of the functions representing these sub-populations.
Wavelet analysis is now frequently used to extract information from epidemiological and other time series. Grenfell et al. \cite{GBK} introduced wavelet analysis for characterizing non-stationary epidemiological time series. 

Cazelles et al. \cite{CCC} use the Morlet wavelets for applications in epidemiology. Lavrova et al. \cite{LPMV} modeled the disease dynamics caused by Mycobacterium tuberculosis in Russia using a sum of two logistic functions (\ref{b1}) (bi-logistic model).

SARS-CoV-2 initially emerged in China, at the end of 2019; after Chinese scientists identified the sequence of the new virus 
\cite{ZYW}, this information was shared with the international community. Since then, a lot of articles were written and published, describing from different points of view, the new SARS-CoV-2 coronavirus and the COVID-19 disease, caused by the virus. We will point out only some of them. Fokas et al. {FDK} used a generalization of the logistic function for forecasting the number of individuals reported to be infected with SARS-CoV-2 in different countries. Krantz et al. \cite{KPR} proposed a two-phase procedure (combining discrete graphs and Meyer wavelets) for constructing true epidemic growth.  A method similar to that one from Lavrova et al. \cite{LPMV} was used by E. Vanucci and L. Vanucci \cite{VV} for predicting the end date of Covid-19 disease in Italy. 

The outline of the present paper is as follows. In Sec.~\ref{sec2}, we discuss the basic properties of Riccati's equation, logistic equation, and logistic curve. For this purpose, we use Eulerian numbers.  Sec.~\ref{sec3} and Sec.~\ref{sec4} are devoted to logistic wavelets. In  Sec.~\ref{sec5} we model the cumulative number of persons reported to be infected by SARS-CoV-2 in the United States as a sum of several logistic functions. 

We use the following convention for the Fourier transform:  
\begin{equation}\label{e1}
	\hat{f}(\xi)=\frac{1}{\sqrt{2\pi}}\int_{-\infty}^{\infty}f(x)e^{-i\xi x} dx,
\end{equation}
where $f\in L^{1}(\mathbb{R})\cap L^{2}(\mathbb{R})$.

\section{Logistic function and its derivatives}\label{sec2}
The logistic equation is defined as
\begin{equation}\label{a1}
	u'(t)=\frac{s}{u_{max}}\:u(u_{max}-u),\quad u(0)=u_{0}.
\end{equation}
where $t$ is time, $u=u(t)$ is the unknown function, $s, u_{max}$ are constants.  The constant $u_{max}$ is called the saturation level. The integral curve $u(t)$ fulfilling condition $0<u(t)<u_{max}$ is known as the logistic function.

After solving (\ref{a1}) we get the logistic function in the following form
\begin{equation}\label{b1}
	u(t)=\frac{u_{max}}{1+e^{-s(t-t_0)}},
\end{equation}
where $t_0$ is its inflection point, which is related to the initial condition $\displaystyle u(0)=u_{0}=\frac{u_{max}}{1+e^{st_0}}$, therefore $\displaystyle t_0=\frac{1}{s}\log\Big(\frac{u_{max}-u_{0}}{u_{0}}\Big)$. 

Equation (\ref{a1}) is a particular case of Riccati's equation with constant coefficients
\begin{equation}\label{c1}
	u'(t)=r(u-u_{1})(u-u_{2}).
\end{equation}
 The constants $r\neq 0,\;u_{1},\;u_{2}$ can be generally real or complex numbers. 

If $u(t)$ is a solution of (\ref{c1}) then it is known a formula for the $n$th derivative $u^{(n)}(t)$ ($n=2,3,4,\ldots$)  of $u(t)$ expressing it as a polynomial of the function $u(t)$ itself:
\begin{equation}\label{d1}
u^{(n)}(t) = r^{n}\sum\limits_{k=0}^{n-1}\Eulerian{ n}{k }
(u-u_{1})^{k+1}(u-u_{2})^{n-k}
\end{equation}
where $n=2,3,\ldots $ and $\displaystyle \Eulerian{ n}{k }$ denotes the Eulerian number (number of permutations of the set $\{1,2,\ldots ,n\}$ having $k,\: (k=0,1,2,\ldots ,n-1)$ permutation ascents, see Graham et al \cite{GKP}). The first few Eulerian numbers are given in the Table~\ref{tab1}. 
\begin{table}
\begin{center}
\begin{tabular}{|c|c c c c c c c c}
\hline
n & $\displaystyle \Eulerian{n }{ 0} $ & $\displaystyle \Eulerian{n}{1} $ & $\displaystyle \Eulerian{n}{2} $ & 
$\displaystyle \Eulerian{n}{3} $ & $\displaystyle \Eulerian{n}{4} $ & $\displaystyle \Eulerian{n}{5} $ & 
$\displaystyle \Eulerian{n}{6} $ & $\displaystyle \Eulerian{n}{7} $\\ \hline
0 & 1 & & & & & & &\\
1 & 1 & 0 & & & & & &\\
2 & 1 & 1 & 0 & & & & &\\
3 & 1 & 4 & 1 & 0  & & & &\\
4 & 1 & 11 & 11 & 1 & 0 & & &\\
5 & 1 & 26 & 66 & 26 & 1 & 0 & &\\
6 & 1 & 57 & 302 & 302 & 57 & 1 & 0 & \\ 
7 & 1 & 120 & 1191 & 2416 & 1191 & 120 & 1 & 0 \\
\hline
\end{tabular}
\end{center}
\caption{Eulerian numbers}
\label{tab1}
\end{table}

Formula (\ref{d1}) was discussed during the Conference ICNAAM 2006
(September 2006) held in Greece and it appeared, with an inductive proof, in paper \cite{Rz1} (see also \cite{Rz2}). Independently the formula has been considered and proved, with the proof based on generating functions, by Franssens \cite{F}.
The polynomial of $u$, of order $(n+1)$,  appearing on the right-hand side of (\ref{a1}) is known in the literature as a kind of the 
so-called derivative polynomials. It is easy to see that all $(n+1)$ roots of the polynomial are simple and lie in the interval $[u_{1}, u_{2}]$.  The derivative polynomials have been recently intensively studied.

Formula (\ref{d1}) applied to the particular case of the logistic equation (\ref{a1}) is as follows:
\begin{equation}\label{a2}
	u^{(n)}(t)= \left(-\frac{s}{u_{max}} \right)^{n}\;\sum\limits_{k=0}^{n-1}\Eulerian{n}{ k}
u^{k+1}(u-u_{max})^{n-k}.
\end{equation}

The polynomial of the variable $u$ and of order $(n+1)$ on the right hand side of (\ref{a2}) is uniform in the sense of the following.
\begin{remark}\label{rem1}
If $u_0$ is a root of the polynomial on the right hand side of (\ref{a2}), i.e.,
\begin{equation}\label{b2}
	\sum\limits_{k=0}^{n-1}\Eulerian{ n}{k }
u_0^{k+1}(u_0-u_{max})^{n-k}=0,
\end{equation}
then dividing both sides of (\ref{b2}) by $u_{max}^{n+1}$ we get
	\[\sum\limits_{k=0}^{n-1}\Eulerian{ n}{k }
\left(\frac{u_0}{u_{max}}\right)^{k+1}\left(\frac{u_0}{u_{max}}-1\right)^{n-k}=0.
\]

Thus $u_0$ is a root of the derivative polynomial on the right hand side of (\ref{a2}) if  $u_0/u_{max}$ is the root of the polynomial
\begin{equation}\label{c2}
	P_{n+1}(u):=(-1)^{n}\sum\limits_{k=0}^{n-1}\Eulerian{ n}{k }
u^{k+1}(u-1)^{n-k}.
\end{equation}
\end{remark}

Let us write down, using formula (\ref{a2}) and the notation of (\ref{c2}), the first few derivatives of the logistic function, which fulfills equation (\ref{a1}). By Remark~\ref{rem1} we can assume, without loss of the generality, that $u_{max}=1$ and $s=1$. 

We obtain successively:
\begin{align*}
	u'(t)=&u(1-u)=-u(u-1)=P_{2}(u),\\
	u''(t)=& u(u-1)^{2}+u^{2}(u-1)=P_{3}(u),\\
	u'''(t)=&-u(u-1)^{3}-4u^{2}(u-1)^{2}-u^{3}(u-1)=P_{4}(u),\\
	u^{(4)}(t)=& u(u-1)^{4}+11u^{2}(u-1)^{3}+11u^{3}(u-1)^{2}+u^{4}(u-1)=P_{5}(u),\\
	u^{(5)}(t)=&-u(u-1)^{5}-26u^{2}(u-1)^{4}-66u^{3}(u-1)^{3}-26u^{4}(u-1)^{2}-u^{5}(u-1)=P_{6}(u).
\end{align*}

All roots of the polynomials $P_{k}(u) \;\textrm{for}\; k=3,4,5,6\;$ can be calculated explicitly, so the polynomials can be factored and we get

\begin{align*}
	P_{3}(u)=& 2u(u-1)\left(u-\frac{1}{2}\right),\\
	P_{4}(u)=&-6u(u-1)\left(u-\frac{1}{2}-\frac{\sqrt{3}}{6}\right)\left(u-\frac{1}{2}+\frac{\sqrt{3}}{6}\right),\\
	P_{5}(u)=&24u(u-1)\left(u-\frac{1}{2}\right)\left(u-\frac{1}{2}-\frac{\sqrt{6}}{6}\right)\left(u-\frac{1}{2}+\frac{\sqrt{6}}{6}\right),\\
	P_{6}(u)=&-120u(u-1)\!\!\left(u-\frac{1}{2}-\frac{\sqrt{30(15-\sqrt{105})}}{60}\right)\!\!\left(u-\frac{1}{2}-\frac{\sqrt{30(15+\sqrt{105})}}{60}\right) \\
	&\left(u-\frac{1}{2}+\frac{\sqrt{30(15-\sqrt{105})}}{60}\right)\!\!\left(u-\frac{1}{2}+\frac{\sqrt{30(15+\sqrt{105})}}{60}\right).
\end{align*}

Therefore the minimal \emph{positive} root of the polynomial
\begin{align}\label{d2}
P_{4}(u)\;\;&\textrm{is}\;\; \frac{1}{2}-\frac{\sqrt{3}}{6}\approx 0.211,\nonumber\\
P_{5}(u)\;\;&\textrm{is}\;\; \frac{1}{2}-\frac{\sqrt{6}}{6}\approx 0.0917,\\
P_{6}(u)\;\;&\textrm{is}\; \; \frac{1}{2}-\frac{\sqrt{30(15+\sqrt{105})}}{60} \approx 0.0413 \nonumber.
\end{align}

Thus by using Remark 1 we see for example that if at a minimal time $t_1$, $u'''(t_1)=0$ ($t_1$ is simultanously a maximum of $u''(t)$) then the value of the logistic function at this point is $u(t_1)=0.211\:u_{max} $. By (\ref{b1}) we have
	\[u(t_1)=\frac{u_{max}}{1+e^{-s(t_1-t_0)}}=0.211\:u_{max},
\]
from which we calculate
\begin{equation}\label{e2}
	t_0-t_1=\frac{1.319}{s}.
\end{equation}

Similar conclusions can be drawn for the smallest zero of the $u^{(4)}(t)$  (polynomial $P_{5}(u)$) or $u^{(5)}(t)$ (polynomial $P_{6}(u)$) using constants (\ref{d2}).

\section{Wavelets based on the second derivative of the logistic function}\label{sec3}
\begin{figure}
	\begin{center}
	 \includegraphics[height=7cm, width=8cm]{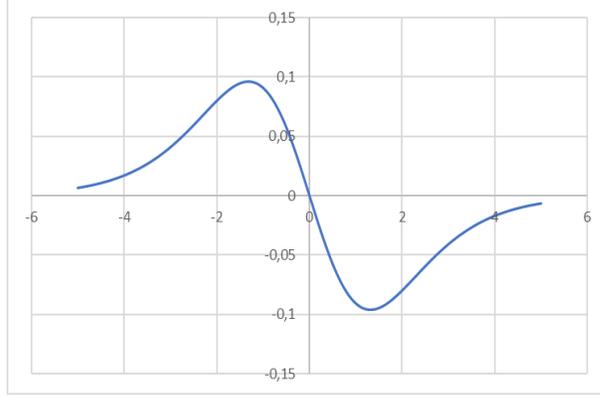}
	\end{center}
	\vspace{-7mm}
	\caption{Wavelet $\psi_2(x)$ }
	\label{fig1}
\end{figure}

Let a wavelet $\psi_2(x)$ (see Figure~\ref{fig1}) be the second derivative of the logistic function $u(x)=\frac{1}{1+e^{-x}}$. Since
 $u'(x)=-u(u-1)$, then by (\ref{d1}) or directly we get
\begin{equation}\label{a4}
	u''(x)=u(1-u)(1-2u), 
\end{equation}
 and by (\ref{a4}) it follows that the wavelet has the following exact form
\begin{equation}\label{b4}
	\psi_2(x)=\frac{1}{1+e^{-x}}\Big(1-\frac{1}{1+e^{-x}}\Big)\Big(1-\frac{2}{1+e^{-x}}\Big)=\frac{e^{-2x}-e^{-x}}{(1+e^{-x})^3}.
\end{equation}

Changing the variable $u=\frac{1}{1+e^{-x}}, u'(x)=u(1-u)$ in the following three integrals we calculate
\begin{align*}
	&\int_{-\infty}^{\infty}\psi_2(x) dx=\int_{0}^{1}(1-2u)du=0,\\
	&\int_{-\infty}^{\infty}|\psi_2(x)| dx=\int_{0}^{1}|1-2u|du=\frac12,\\
	&\int_{-\infty}^{\infty}(\psi_2(x))^2 dx=\int_{0}^{1}u(1-u)(1-2u)^2 du=\frac{1}{30},
\end{align*}
which proves that $\psi_2(x) \in L^{1}(\mathbb{R})\cap L^{2}(\mathbb{R})$.  In fact $\psi_2(x) \in S(\mathbb{R})$ (the space of rapidly decreasing functions on $\mathbb{R}$). We will discuss this in the next section.

By $\psi_2^{+}(x)$ we denote the positive part of $\psi_2(x)$, i.e.,
	\[\psi_2^{+}(x)=\frac12 (\psi_2(x)+|\psi_2(x)|).
\]
Obviously
\begin{equation}\label{bb4}
	\int_{-\infty}^{\infty}(\psi_2^{+}(x))^2 dx=\frac12 \int_{-\infty}^{\infty}(\psi_2(x))^2 dx=\frac{1}{60}.
\end{equation}

The Fourier transform of $\psi_2(x)$ is as follows:
\begin{equation}\label{c4}
	\hat{\psi_2}(\xi)=\frac{1}{\sqrt{2\pi}}\int_{-\infty}^{\infty}\psi_2(x)e^{-i\xi x} dx=\sqrt{\frac{\pi}{2}}\frac{i\xi^2}{\sinh(\pi\xi)}.
\end{equation}

It is well known (see \cite{D}) that a wavelet $\psi(x)\in L^{1}(\mathbb{R})\cap L^{2}(\mathbb{R})$ should satisfy the following
admissibility condition
 \begin{equation}\label{f1}
	2\pi\int_{-\infty}^{\infty}|\xi|^{-1}|\hat{\psi}(\xi)|^2 d\xi < \infty.
\end{equation}

We will show that for $\psi_2(x)$ the condition (\ref{f1}) is satisfied and even the integral can be expressed in a closed form in terms of the Riemann zeta function. Namely, using (\ref{c4}) and the following formula from Dwight's Tables \cite{Dw} (item no $860.519$):
	\begin{equation}\label{d4}
		\int_{0}^{\infty}\frac{x^{p}}{(\sinh (ax))^2} dx=\frac{\Gamma (p+1)}{2^{p-1}a^{p+1}}\zeta(p), \quad a>0,\; p>1,
	\end{equation}

we have
\begin{equation}\label{e4}
	2\pi\int_{-\infty}^{\infty}|\xi|^{-1}|\hat{\psi_2}(\xi)|^2 d\xi = \pi^2\int_{-\infty}^{\infty}\frac{|\xi|^3}{(\sinh(\pi\xi))^2}d\xi=\frac{3\zeta(3)}{\pi^2}.
\end{equation}

We generate a doubly-indexed family of wavelets from $\psi_2$ by dilating and translating,
	\[\psi_2^{a,b}(x)=\frac{1}{\sqrt{a}}\psi_2\Big(\frac{x-b}{a}\Big),
\]
where $a,b\in \mathbb{R}, \; a>0$ and denote by $\psi_2^{+\;a,b}(x)$ the positive part of $\psi_2^{a,b}(x)$.

\section{Wavelets based on higher derivatives of the logistic function}\label{sec4}
\begin{figure}
	\begin{center}
	 \includegraphics[height=7cm, width=8cm]{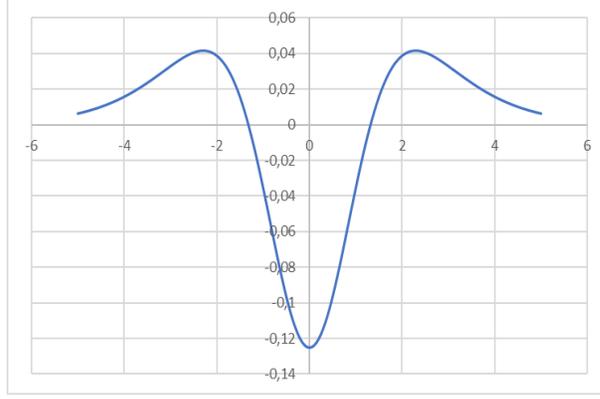}
	\end{center}
	\vspace{-7mm}
	\caption{Wavelet $\psi_3(x)$ }
	\label{fig2}
\end{figure}

Similarly as in the previous section we define a wavelet $\psi_n(x)$ to be the $n$th ($n=3,4,...$) derivative of the logistic function $u(x)=\frac{1}{1+e^{-x}}$. Figure~\ref{fig2} shows graph of the wavelet $\psi_3(x)$. Thus (\ref{d1}) gives
	\[u^{(n)}(x)=(-1)^{n}\sum\limits_{k=0}^{n-1}\Eulerian{ n}{k }u^{k+1}(u-1)^{n-k},
\]
and then $\psi_n(x)$ can be explicitly expressed as
\begin{equation}\label{a5}
	\psi_n(x)=(-1)^{n}\sum\limits_{k=0}^{n-1}\Eulerian{ n}{k }\Big(\frac{1}{1+e^{-x}} \Big)^{k+1}\Big(\frac{1}{1+e^{-x}}-1\Big)^{n-k}=
	\frac{(-1)^{n}\sum\limits_{k=0}^{n-1}\Eulerian{ n}{k }(-e^{-x})^{n-k}}{(1+e^{-x})^{n+1}}.
\end{equation}

By definition, the function $\psi_n(x)$ is an even function for odd n and an odd function when n is even. The numerator of the expression (\ref{a5}) is a polynomial of degree $n$ of the variable $e^{-x}$, while the denominator of degree $n+1$. Therefore for any polynomial $p(x)$ we have $\lim\limits_{x\rightarrow -\infty} p(x)\psi_n(x)=0$. Since $\psi_n(x)$ has the symmetry property then also 
$\lim\limits_{x\rightarrow \infty} p(x)\psi_n(x)=0$. The last conclusion can also be drawn from multiplying the numerator and the denominator of (\ref{a5}) by $e^{(n+1)x}$. From this and from the fact that $\psi_{k+1}(x)=\psi_{k}'(x)$ for any integer $k\ge 2$ it follows that $\psi_{n}(x) \in S(\mathbb{R})$, ($n=2,3,\ldots$).

By (\ref{c4}) we have
\begin{equation}\label{b5}
	\hat{\psi_n}(\xi)=\frac{1}{\sqrt{2\pi}}\int_{-\infty}^{\infty}\psi_n(x)e^{-i\xi x} dx=\sqrt{\frac{\pi}{2}}\frac{(i\xi)^{n-1}\xi}{\sinh(\pi\xi)}.
\end{equation}

Now using (\ref{b5})  and once again formula (\ref{d4}) we can calculate the integral of the admissibility condition (\ref{f1}) as follows:
\begin{equation}\label{c5}
	2\pi\int_{-\infty}^{\infty}|\xi|^{-1}|\hat{\psi_n}(\xi)|^2 d\xi = \pi^2\int_{-\infty}^{\infty}\frac{|\xi|^{2n-1}}{(\sinh(\pi\xi))^2}d\xi=\pi^2 \frac{2\Gamma(2n)}{2^{2n-2}\pi^{2n}}\zeta(2n-1)=\frac{(2n-1)!}{2^{2n-3}\pi^{2n-2}}\zeta(2n-1).
\end{equation}

As usually we generate a doubly-indexed family of wavelets from $\psi_n$ by dilating and translating,
	\[\psi_n^{a,b}(x)=\frac{1}{\sqrt{a}}\psi_n\Big(\frac{x-b}{a}\Big),
\]
where $a,b\in \mathbb{R}, \; a>0, \; n=2,3,\ldots$.

\section{Applications for modeling of SARS-CoV-2 virus epidemics }\label{sec5}

Denote by $y_n^{*}$ total cumulative number of individuals reported to be infected up to $n$th day in a country or a region and by $y_n$
the $7$-day central moving arithmetic average for the sequence $y_n^{*}$, i.e., 
	\[y_n=\frac17\sum\limits_{i=-3}^{3}y_{n+i}^{*}.
\]

We will look, in the sequence $(y_n)$, for points corresponding to the zeros of the second or the third derivative of the logistic function. This is equivalent to detect the points, where the sequence of second differences, 
	\[\Delta^2y_n=y_{n+1}-2y_n+y_{n-1},
\]
takes a value close to zero or a maximum respectively. We will find these points either directly by observing the sequence of second differences $(\Delta^2y_n)$ or detect them by using the wavelet $\psi_2(x)$ and its positive part $\psi_2^{+}(x)$. From the considerations in Sec.~\ref{sec2} and from (\ref{e2}) it follows that parameter $b$ should be determined as that point where the sequence $(\Delta^2y_n)$ changes sign. Parameter $a$ should be chosen in such a way that the distance between the zero and the maximum of $(\Delta^2y_n)$ was approximately $1.319a$. Thus, we obtain two parameters defining the first logistic function (first wave) approximating the time series $(y_n)$. It remains to determine the third parameter of the first wave, i.e., its saturation level $y_{max}$. Assuming that $(y_n)$ initially follows a logistic function $\displaystyle y_n\approx y(n)=\frac{y_{max}}{1+\exp(-\frac{n-b}{a})}$ and since by definition it holds 
	\[ y''(x)=\frac{y_{max}}{a^{3/2}}\psi_2^{a,b}(x), 
\]
then by (\ref{bb4}) we get successively

\begin{align}
	\sum\limits_{n}&\Delta^2y_n\psi_2^{+\;a,b}(n)\approx \sum\limits_{n}\Delta^2y(n)\psi_2^{+\;a,b}(n)
	\approx \int_{-\infty}^{\infty}y''(x)\psi_2^{+\;a,b}(x)dx
	 =\int_{-\infty}^{\infty} \frac{y_{max}}{a^{3/2}}\psi_2^{a,b}(x)\psi_2^{+\;a,b}(x)dx \nonumber\\
	&=\frac{y_{max}}{a^{3/2}}
	\int_{-\infty}^{b}(\psi_2^{+\;a,b}(x))^2 dx= \frac{y_{max}}{60a^{3/2}}.\label{a6}
\end{align}

Using (\ref{a6}) we can estimate $y_{max}$ as follows
\begin{equation}\label{b6}
y_{max}\approx 60a^{3/2}\sum\limits_{n}\Delta^2y_n\psi_2^{+\;a,b}(n),
\end{equation}
	
Parameters $a$ and $b$ can also be estimated by maximizing locally the integral on the left-hand side of (\ref{a6}). Thus we find in the sequence $\Delta^2y_n$ the best pattern corresponding to the positive part of the wavelet $\psi_2^{a,b}$. To avoid the situation that the next wave, immediately following the previous one, could distort our findings we use here the positive part $\psi_2^{+\;a,b}$, not the whole wavelet  $\psi_2^{a,b}$. The saturation level of the first wave can also be estimated as twice the value of the sequence $(y_n)$ at the point where $(\Delta^2y_n)$ changes signs (inflection point) or  its maximal value multiplied by $1/0.211$ (zero of the third derivative).	Having found the values of the parameters $a$, $b$, and $y_{max}$ for the first wave, we create a new time series by subtracting the first wave from $y_n$, i.e.,
	\[z_n=y_n-\frac{y_{max}}{1+\exp(-\frac{n-b}{a})},
\]
and with the sequence $z_n$ we proceed in the same way as with $y_n$, calculating successive logistic waves. After this we use the nonlinear Generalized Reduced Gradient method to optimize the values of saturation levels (but not $a$'s and $b$'s). All data were collected from the https://www.worldometers.info/coronavirus/ platform.

\subsection{The spread of infections in the United States}
Let us use the theory to build a model for the total cumulative number of individuals reported to be infected by SARS-CoV-2 in the USA.  We assumed the observation period of 189 days, from March 13 ($n=1$, the first day when the number of cases exceeded $2,000$) to September 17, 2020 ($n=189$). All calculations were performed in Excel. 

Using the above-described procedure we have got the approximating function as a sum of the following waves
\begin{equation}\label{c6}
	f(x)=\frac{630,913}{1+\exp(-\frac{x-25}{6.6})}+\frac{1,085,184}{1+\exp(-\frac{x-52}{9})}+\frac{3,288,916}{1+\exp(-\frac{x-126}{14.6})}+\frac{2,846,457}{1+\exp(-\frac{x-174}{23.3})}.
\end{equation}

\begin{figure}
	\begin{center}
	 \includegraphics[height=5cm, width=8cm]{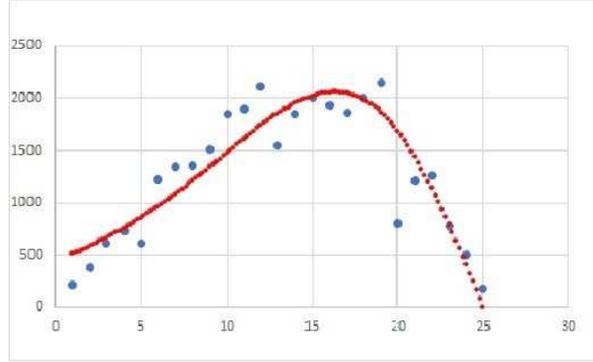}
	\end{center}
	\vspace{-7mm}
	\caption{Scaled first wave's wavelet $\psi_2^{+\;a,b}$, $a=25, b=6.6$ fitted to the data}
	\label{fig3}
\end{figure}

\begin{figure}
	\begin{center}
	 \includegraphics[height=5cm, width=8cm]{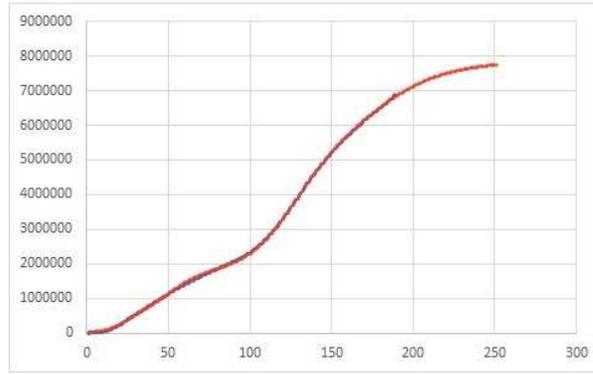}
	\end{center}
	\caption{Approximation of the total number of infections in the USA , blue points - observed values, red line - approximating function $f(x)$ (\ref{c6}) }
	\label{fig4}
\end{figure}

Figure~\ref{fig3} shows the positive part of the wavelet (scaled) $\psi_2^{+\;a,b}$, $a=25, b=6.6$ fitted to the second differences $(\Delta^2y_n)$. Figure~\ref{fig4} shows the total cumulative number of individuals reported to be infected by SARS-CoV-2 in the USA in the period: March 13 ($n=1$) - September 17 ($n=189$), (blue points) and the approximating function $f(x)$ (\ref{c6}), (red points). 

The MAD (Mean Absolute Deviation) value of the approximation equals
	\[\textrm{MAD}=\frac{1}{189}\sum\limits_{n=1}^{189}|y_n-f(n)|=20,882.53.
\]

 We have  calculated SMAPE (Symmetric Mean Absolute Percent Error) value  in two versions:
\begin{align*}
	\textrm{SMAPE}_1&=\frac{2}{189}\sum\limits_{n=1}^{189}\frac{|y_n-f(n)|}{y_n+f(n)}=0.08217,\\
	\textrm{SMAPE}_2&=\frac{\sum\limits_{n=1}^{189}|y_n-f(n)|}{\sum\limits_{n=1}^{189}(y_n+f(n))}=0.003679.\\
\end{align*}
 
The error SMAPE$_1$ strongly depends on the initial values of the time series. If we omit the first 19 observations, then the error calculated for the shorter period of $170$ days from April 1 ($n=20$) to September 17 reduces to
	\[\textrm{SMAPE}_1=\frac{2}{170}\sum\limits_{n=20}^{189}\frac{|y_n-f(n)|}{y_n+f(n)}=0.01195.
\]

\subsection{The spread of infections in the United Kingdom}
\begin{figure}
	\begin{center}
	 \includegraphics[height=5cm, width=8cm]{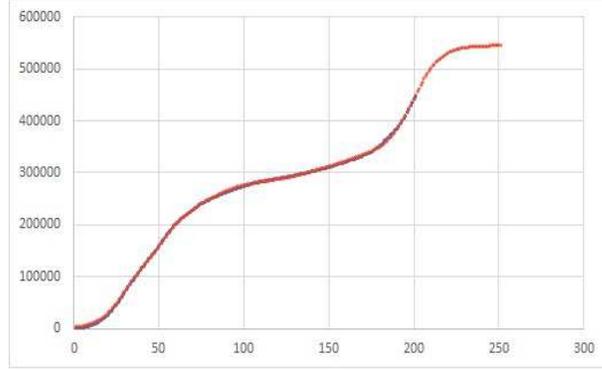}
	\end{center}
	\caption{Approximation of the total number of infections in the UK, blue points - observed values, red line - approximating function $f(x)$ (\ref{d6}) }
	\label{fig5}
\end{figure}

The same procedure applied to the total cumulative number of individuals reported to be infected by SARS-CoV-2 in the United Kingdom for the period of $201$ days from March 13 ($n=1$) to September 29 ($n=201$) gives the following approximating function (see Figure~\ref{fig5})
\begin{align}\label{d6}
	f(x)=&\frac{130,000}{1+\exp(-\frac{x-29}{7.4})}+\frac{83,595}{1+\exp(-\frac{x-52}{6.1})}+\frac{30,441}{1+\exp(-\frac{x-68}{5.6})}+\frac{29,688}{1+\exp(-\frac{x-86}{6.2})}+\frac{3,405}{1+\exp(-\frac{x-102}{4.7})} \nonumber\\
	&+\frac{74,547}{1+\exp(-\frac{x-152}{18.7})} +\frac{194,489}{1+\exp(-\frac{x-200}{7.6})}.
\end{align}

The approximation errors are: $\textrm{MAD}=1,443.37$, $\textrm{SMAPE}_1=0.05955$ (for the shorter period of $182$ days, beginning from April 1, $\textrm{SMAPE}_1=0.007123$), $\textrm{SMAPE}_2=0.003047$.

\section{Conclusions and further work}
In the paper, we have proved, by using the Eulerian numbers, properties of the logistic function related to zeros of its successive derivatives. We also used logistic wavelets to estimate the parameters of a logistic curve that best fits to a given time series with a potential logistic trend. Then we described, based on the data from the United States and the United Kingdom, that the total reported number of SARS-CoV-2 infections can be modeled, in a natural way, as a sum of several logistic functions. The theory and the procedure can be applied to model the number of infections in any country or a region.

In our further work, we intend to use, in a similar way, the logistic wavelets of higher order (see Sec.~\ref{sec4}). Using some appropriate special numbers we are going to define analogous wavelets for the Gompertz function (see some initial calculations \cite{Rz3}, \cite{Rz4}) or for the fractional logistic functions (some preliminary theorems see \cite{Rz5}). 

\vspace{7mm}

{\large\textbf{Funding statement}}

This research was partially funded by the 'IDUB against COVID-19' project granted by the Warsaw University of Technology (Warsaw, Poland) under the program Excellence Initiative: Research University (IDUB).


\begin{thebibliography}{10}

\bibitem{CCC} B. Cazelles, K. Cazelles and M. Chavez, Wavelet analysis in ecology and epidemiology: impact of statistical tests. \emph{J. R. Soc. Interface} \textbf{11} (2014): 20130585. http://dx.doi.org/10.1098/rsif.2013.0585
\bibitem{D} I. Daubechies, \emph{Ten lectures on wavelets}, 2nd ed., Philadelphia: SIAM, 1992. CBMS-NSF regional conference series in applied mathematics 61
\bibitem{Dw} H. B. Dwight, \emph{Tables of integrals and other mathematical data}, 4th ed., The Macmillan Company, New York, 1961.
\bibitem{FDK} A. S. Fokas, N. Dikaios and G. A. Kastis, Mathematical models and deep learning for predicting the number of individuals reported to be infected with SARS-CoV-2, \emph{J. R. Soc. Interface} \textbf{17} (2020): 20200494. http://dx.doi.org/10.1098/rsif.2020.0494
\bibitem{F} G. R. Franssens, Functions with derivatives given by polynomials in the function itself or a related function, \emph{Analysis Mathematica} \textbf{33} (2007), 17--36.
\bibitem{GKP} R. L. Graham, D. E. Knuth and O. Patashnik, \emph{Concrete Mathematics: A Foundation for Computer Science}, Reading MA: Addison Wesley, 1994.
\bibitem{GBK} B.T. Grenfell, O. N. Bj{\o}rnstad and J. Kappey, Travelling waves and spatial hierarchies in measles epidemics. \emph{Nature} \textbf{414} (2001), 716–-723. https://doi.org/10.1038/414716a
\bibitem{KM} W. O. Kermack, A. G. McKendrick, A contribution to the mathematical theory of epidemics, \emph{Proc. R. Soc. Lond. A} \textbf{115} (1927), 700–-721. https://doi.org/10.1098/rspa.1927.0118
\bibitem{KPR} S. G. Krantz, P. Polyakov and A.S.R.S. Rao, True epidemic growth construction through harmonic analysis, \emph{J. Theor. Biol.} \textbf{494} (2020): 110243. https://doi.org/10.1016/j.jtbi.2020.110243
\bibitem{LPMV} A. I. Lavrova, E. B. Postnikov, O. A. Manicheva and B. I. Vishnevsky, Bi-logistic model for disease dynamics caused
by Mycobacterium tuberculosis in Russia, \emph{R. Soc. Open Sci.} \textbf{4} (2017): 171033. https://doi.org/10.1098/rsos.171033
\bibitem{Rz1} G. Rz\c{a}dkowski, Eulerian numbers and Riccati's differential equation, (Eds. T.E. Simos) Proceedings of  ICNAAM 2006, Wiley--VCH Verlag (2006), 291--294.
\bibitem{Rz2} G. Rz\c{a}dkowski, Derivatives and Eulerian numbers, \emph{Amer. Math. Monthly} \textbf{115} (2008), 458--460. 
\bibitem{Rz3} G. Rz\c{a}dkowski, W. Rz\c{a}dkowski, P. Wójcicki, On some connections between the Gompertz function and special numbers, \emph{J. Nonlinear Math. Phys.} \textbf{3} (2015), 374–-380. http://dx.doi.org/10.1080/14029251.2015.1079419 
\bibitem{Rz4} G. Rz\c{a}dkowski, I. Głażewska, K. Sawińska, The Gompertz function and its applications in management, \emph{Foundations of Management} \textbf{7} (2015), 185--190. DOI: 10.1515/fman-2015-0035
\bibitem{Rz5} G. Rz\c{a}dkowski, M. Urlińska, Some applications of the generalized Eulerian numbers, \emph{J. Comb. Theory Ser. A.} \textbf{163} (2019), 85--97. DOI: https://doi.org/10.1016/j.jcta.2018.11.012
\bibitem{VV} E. Vanucci, L. Vanucci, Forecast Covid-19 end date in Italy by logistics waves, https://www.researchgate.net/publication/341104205 (Access: 2020 August 21).
\bibitem{ZYW} Zhou, P., Yang, X., Wang, X. et al. A pneumonia outbreak associated with a new coronavirus of probable bat origin. \emph{Nature} \textbf{579} (2020), 270–-273. https://doi.org/10.1038/s41586-020-2012-7
\end{thebibliography}
\end{document}